\documentclass[aps,prl,preprint,superscriptaddress]{revtex4-1}
\usepackage[dvips]{graphicx}
\usepackage{comment}
\usepackage{color}
\usepackage{txfonts}
\usepackage{bm}
\usepackage{here}

\begin{document}
\def\val{3}

\title{Magnetic-field and pressure phase diagram of the triangular lattice antiferromagnet CsCuCl$_3$ clarified by magnetic susceptibility measured with a proximity detector oscillator }


\author{K. Nihongi}
\author{T. Kida}
\author{Y. Narumi}
\affiliation{Center for Advanced High Magnetic Field Science (AHMF), Graduate School of Science, Osaka University, Toyonaka, Osaka 560-0043, Japan}

\author{J. Zaccaro}
\affiliation{Grenoble Alpes University, CNRS, Grenoble INP, Institut N$\acute{e}$el, 38042 Grenoble, France}

\author{Y. Kousaka}
\affiliation{Department of Physics and Electronics, Osaka Prefecture University, Sakai, Osaka 599-8531, Japan}

\author{K. Inoue}
\affiliation{Graduate School of Science, Hiroshima University, Hiroshima 739-8526, Japan}
\affiliation{Center for Chiral Science, Hiroshima University, Hiroshima 739-8526, Japan}

\author{K. Kindo}
\author{Y. Uwatoko}
\affiliation{The Institute for Solid State Physics, The University of Tokyo, Kashiwa, Chiba 277-8581, Japan}

\author{M. Hagiwara} \thanks{Corresponding author}
\affiliation{Center for Advanced High Magnetic Field Science (AHMF), Graduate School of Science, Osaka University, Toyonaka, Osaka 560-0043, Japan}
\affiliation{Center for Chiral Science, Hiroshima University, Hiroshima 739-8526, Japan}
\email[e-mail:]{hagiwara@ahmf.sci.osaka-u.ac.jp}



\date{\today}

\begin{abstract}
The effect of pressure ($P$) on magnetic susceptibility of CsCuCl$_3$ was examined in magnetic fields ($\rm \mu_0$$H$) of up to 51 T using a proximity detector oscillator (PDO), and the $H$-$P$ phase diagram of CsCuCl$_3$ was constructed over the saturation field ($H_{\rm sat}$). We found that, with increasing $P$, $H_{\rm sat}$ increases and the uud-phase that appeared at $P$ = 0.7 GPa widened. Based on comparison between the experimental and calculated $H$-$P$ phase diagrams, the Y-phase was predicted to appear above 1.7 GPa.   The interchain antiferromagnetic exchange interaction in the $ab$-plane was evaluated and found to increase with increasing $P$, which is consistent with a previous study under high pressure [D. Yamamoto {\it et al.}, Nat. Commun. {\bf 12}, 4263 (2021).].  Moreover, an anomaly was observed below $P$ = 0.6 GPa just below $H_{\rm sat}$ and might be a new phase transition derived from nonlinear response caused by the PDO technique.
\end{abstract}


\maketitle

\section{Introduction}
Geometrically frustrated magnets have been studied over the past several decades owing to a rich variety of exotic physical phenomena \cite{ Anderson.spinliquid,Balents.spinliquid}. One of the most typical geometrically frustrated magnets is the two-dimensional triangular lattice antiferromagnet (TLA) \cite{starykh2015unusual}. The ground state infinitely degenerates in magnetic fields, and a small perturbation, such as easy-plane anisotropy, lifts the degeneracy and stabilizes the one-energy state \cite{RN68}, for example, an umbrella spin structure in low-magnetic fields normal to the easy plane. For classical spin systems, extensive theoretical studies have been conducted \cite{Kawamura.Miyashita, Chubukov, seabra2011phase}, and one of the experimental realizations is RbFe(MoO$_4$)$_2$ \cite{Inami, SmirnovPRB.75.13.2007}. When spins are small, quantum fluctuations play an important role in lifting degeneracy in some cases. For TLAs, the up-up-down (uud) state, which corresponds to a magnetization plateau at one-third of the saturation magnetization, is an exotic ground state \cite{Kawamura.Miyashita, Chubukov, seabra2011phase, Sakai, ColettaPRB.94.7.2016} caused by the order-by-disorder mechanism \cite{henley1989ordering} in magnetic fields. To date, Cs$_2$CuBr$_4$ \cite{tanaka2002magnetic, OnoPRB2003, OnoJPCM2004} (not equilateral) and Ba$_3$CoSb$_2$O$_9$ \cite{ShirataPRL.108.057205.2012, SusukiPRL.110.267201.2013}  are known to exhibit the uud ground state of $\it S$ = 1/2 TLAs in magnetic fields. \par

CsCuCl$_3$ has been studied for a long time as one of ABX$_3$ compounds \cite{AchiwaJPSJ275611969}. It is a TLA compound formed by ferromagnetic chains and exhibits a quantum phase transition in magnetic fields \cite{Nojiri, Nikuni}. The crystal structure of CsCuCl$_3$ at room temperature is hexagonal with a space group $\it P$6$_1$22 or $\it P$6$_5$22 \cite{wells1947332, schlueter1966redetermination}. Magnetic Cu$^{2+}$ ions of CsCuCl$_3$ with $\it S$ = 1/2 form a helical chain structure with a sixfold periodicity along the $c$-direction and triangular lattices in the $ab$-plane. The spin Hamiltonian for CsCuCl$_3$ in magnetic fields \cite{Nikuni} is given by, 

\begin{eqnarray}
\mathcal{H}  =&-&2J_0\sum_{i,n}( \mbox{\boldmath $S$}_{i,n} \cdot  \mbox{\boldmath $S$}_{i,n+1} + \eta (S_{i,n}^xS_{i,n+1}^x + S_{i,n}^yS_{i,n+1}^y)) \nonumber\\& + &2J_1\sum_{ <i,j>,n}( \mbox{\boldmath $S$}_{i,n} \cdot  \mbox{\boldmath $S$}_{i,n+1})
 - \sum_{ i,n} \mbox{\boldmath $D$}_{n,n+1}\cdot( \mbox{\boldmath $S$}_{i,n} \times  \mbox{\boldmath $S$}_{i,n+1}) \nonumber\\ & - &g{\rm \mu_B} H\sum_{ i,n}S_{i,n}^z,
\end{eqnarray}

\noindent
where $S_{i,n}$ represents a spin located at the $i$-th site in the $n$-th $ab$-plane, and the summation $\langle$$ij$$\rangle_n$ is taken over all nearest-neighbor pairs in the $n$-th $ab$-plane. The $z$-axis is assumed to be parallel to the $c$-axis. The first and second terms are the intrachain ferromagnetic exchange (exchange constant $J_0 > 0$) and interchain antiferromagnetic exchange (exchange constant $J_1 > 0$)  interactions, respectively.  $\eta$ ($>$ 0) represent a weak anisotropic exchange ratio for the easy plane. The third term represents the Dzyaloshinskii-Moriya (DM) interaction where $\mbox{\boldmath $D$}$$_{\rm n,n+1}$ represents the DM vector between adjacent spins in the $n$-th and ($n$+1)-th $c$-planes. The fourth term is the Zeeman energy. $g$ and $\rm \mu_{\rm B}$ are the $g$-factors for the $z$-axis and Bohr magneton, respectively. \par
The intrachain and interchain exchange constants were evaluated based on the following analyses: $J_0$/k$_B$ = 24 $\pm$ 3 K and $J_1$/k$_B$ = 3.8 $\pm$ 1 K from magnetic susceptibilities \cite{Tazuke}, $J_0$/k$_B$ = 28 K and $J_1$/k$_B$ = 4.9 K from electron spin resonance (ESR) excitation modes for $H \parallel c$ \cite{Tanaka.ESR.CsCuCl3},  and $J_0$/k$_B$ = 28 K (2.4 meV) and $J_1$/k$_B$ = 4.6 K (0.4 meV) from inelastic neutron scattering data \cite{mekata1995magnetic}. The Cu spins exhibited a 120$^{\circ}$ spin structure in the $ab$-plane below the N$\acute{e}$el temperature $T_{\rm N}$ = 10.7 K \cite{AdachiJPSJ49.543.1980, Weber}. The Cu chain structure along the $c$-axis allows DM interaction with a uniform $\bm D$ vector pointing in the chain direction ($D$/k$_B$ = 5.1 K \cite{Tanaka.ESR.CsCuCl3} or 5.8 K \cite{mekata1995magnetic}) owing to the chiral crystal structure \cite{AdachiJPSJ49.543.1980, Kousaka}. Below $T_{\rm N}$, the nearest-neighbor spins along the $c$-axis were rotated with a pitch angle of $\theta$ =  5.1 $\pm$0.1$^\circ$ in the $ab$-plane \cite{AdachiJPSJ49.543.1980}. In the following analysis of our experimental data, we used the parameter values obtained by the analysis of ESR excitation modes \cite{Tanaka.ESR.CsCuCl3}: $J_0$/k$_B$ = 28 K and $J_1$/k$_B$ = 4.9 K, and $D$/k$_B$ = 5.1 K . \par
The magnetization ($M$) of CsCuCl$_3$ for $H \parallel c$ at 1.1 K increased almost linearly and showed a jump at 12.5 T, then increased again and saturated at 31 T \cite{Nojiri}. Nikuni and Shiba interpreted the magnetization process of CsCuCl$_3$ for $H \parallel c$  \cite{Nikuni} using the spin Hamiltonian (Eq.(1)).  They eliminated the asymmetric part from the Hamiltonian by rotating the $xy$-plane with pitch $q$ = tan$^{-1}$ ($D$/2$J_0$ (1 + $\eta$)).  Thereafter, the spin Hamiltonian can be simply rewritten as:

\begin{eqnarray}
\mathcal{H}  = &-&\sum_{ i,n}[2\tilde{J}_0  (S_{i,n}^xS_{i,n+1}^x + S_{i,n}^yS_{i,n+1}^y) + 2J_0S_{in}^zS_{in+1}^z]\nonumber\\&+ &2J_1\sum_{<ij>n} \mbox{\boldmath $S$}_{in} \cdot  \mbox{\boldmath $S$}_{jn} - g{\rm \mu_B} H\sum_{ i,n}S_{i,n}^z,
\end{eqnarray}

\noindent
where $\tilde{J}_0$ = $J_0$$\sqrt{(1 + \eta)^2 + (D/2J_0)^2}$. They defined the easy-plane anisotropy as,

\begin{equation}
\Delta \equiv (\tilde{J}_0 - J_0)/J_1.
\end{equation}

\noindent
Thereafter, the value of saturation field is given by,

\begin{equation}
H_{\rm sat} = (18 + 4\Delta)J_1S/g\rm \mu_B.
\end{equation}

\noindent
They interpreted that the jump in $M$ of CsCuCl$_3$ for $H$ $\parallel$ $c$ was caused by the quantum phase transition from the umbrella to 2-1 coplanar phase. Their theoretical predictions were confirmed by neutron diffraction \cite{Mino} in magnetic fields.  \par

The application of high pressure is a powerful method that can alter the magnetic properties of a material by shrinking its crystal lattice. As for TLAs, magnetic anisotropy and the exchange interaction between magnetic ions vary through the change in distance and angle between the neighboring magnetic ions through intervening nonmagnetic ions by applying pressure. For the anisotropic TLA Cs$_2$CuCl$_4$, the appearance of a number of field-induced magnetic phase transitions upon pressure application have been reported, accompanied by an increase in the exchange coupling ratio $J^{\prime}$/$J$, where $J$ and $J^{\prime}$ are the exchange interactions on the horizontal and diagonal bonds in the spatially anisotropic triangular lattice, respectively \cite{Zvyagin}.   \par

Sera {\it et al.} \cite{Sera} performed magnetization measurements of CsCuCl$_3$ at 1.4 K in magnetic fields ($\rm \mu$$_0$$H$ up to 15 T) along the $c$-axis under high pressure ($P$ up to 0.9 GPa) and observed the magnetization plateau with 0.34 $\rm \mu$$_B$/Cu$^{2+}$ under $\it P$ $\geq$ 0.68 GPa, which corresponds to the uud phase in $M$. Hosoi {\it et al.} tried to explain the magnetization under high pressure by assuming that $\Delta$ was sensitive to pressure \cite{Hosoi}. They constructed the magnetic phase diagram of CsCuCl$_3$ at 0 K under pressure by calculating the spin Hamiltonian based on spin-wave theory with $\Delta$ ($\Delta$ = 0.07 at ambient pressure) as a variable of pressure. $\Delta$ decreased with increasing pressure, and the uud and Y coplanar phases under high pressure ($\Delta$ $\leq$ 0.06) were predicted to appear with the fixed parameters of $J_0$ and $\tilde{J}_0$. Yamamoto {\it et al.} demonstrated theoretically that the ratio of $J_0$ to $J_1$ reduces with pressure, and then the uud and Y coplanar phases appeared under $P \geq$ 0.75 GPa, corresponding to the results of magnetic measurement of CsCuCl$_3$ under high pressure up to 1.21 GPa in magnetic fields of up to 5 T \cite{Yamamoto.Nature.12.4263.2021}. In both theories, the change in the ratio of $J_0$ to $J_1$ was important for the appearance of the field- and pressure-induced magnetic phases. 
A quantitative comparison of the $H$-$P$ phase diagram of CsCuCl$_3$ requires $J_1$ under high pressure which was directly estimated from $H_{\rm sat}$. Therefore, in this study, we investigated the pressure dependence of the entire magnetization process of CsCuCl$_3$ over the saturation field using a proximity detector oscillator (PDO) system. \par
 The remainder of this study is organized as follows. In Section 2, the sample preparation of single crystals of CsCuCl$_3$, experimental methods and conditions are described. The experimental results of the pressure dependence of magnetic susceptibility up to 1.72 GPa and the magnetic-field ($H$) versus pressure ($P$) phase diagram obtained experimentally are presented in Section 3. In Section 4, the constructed $H$-$P$ phase diagram is compared with the calculated phase diagram by Hosoi {\it et al.} \cite{Hosoi}. The similarities and differences between them are discussed. Finally, the origin of the anomaly observed immediately below the saturation field in the $c$-direction is discussed, followed by a summary of this study. 

 \section{Experimental}
 
 Single crystals of CsCuCl$_3$ were grown by evaporating an aqueous solution containing equimolar amounts of CsCl and CuCl$_2$$\cdot$2H$_2$O. Pulsed magnetic fields of up to 51 T with a pulse duration of 35 ms were generated using a nondestructive pulse magnet at AHMF at Osaka University. High-field magnetization measurements were performed using an induction method with a pick-up coil at temperatures down to 1.4 K at ambient pressure. A PDO system \cite{Altarawneh} was utilized to measure the change in the PDO resonance frequency ($\Delta$$f$), which corresponded to the field change of the magnetization ($\Delta$$M$/$\Delta$$H$), namely magnetic susceptibility,  in a magnetic insulator, as explained below, under high pressures of up to 1.72 GPa. When the magnetic insulator was placed inside the sensor coil of the LC tank circuit, the change in $\Delta$$f$ in the magnetic field was proportional to $\Delta$$M$/$\Delta$$H$ \cite{Ghannadzadeh}. In this study, we used a NiCrAl piston cylinder-type pressure cell (PCC) with the inner and outer diameters of 2.0 and 8.6 mm, respectively, and the pressure medium Daphne 7373 oil (Idemitsu Kosan Co.,Ltd.). The pressure in the sample space was calibrated using the change in the superconducting transition temperature of Sn \cite{smith1967will}. A large field sweep rate of pulsed magnetic fields induced the Joule heating caused by eddy currents generated in a conducting material. Based on our preliminary experiments, we found that the temperature in the sample space changed slightly until approximately 6.5 ms (approximately 40 T in the ascending process when the maximum field was 51 T) from the start of magnetic-field generation.   

 \section{Experimental Results}

    \begin{figure*}[t]
\includegraphics[keepaspectratio, scale=0.9]{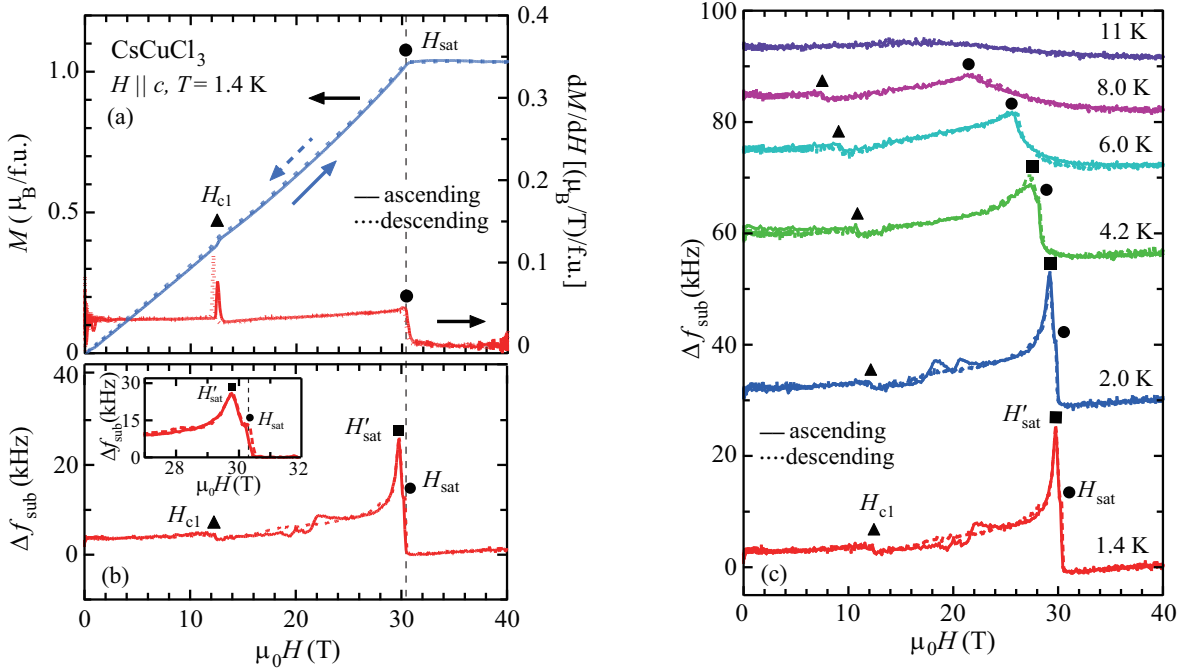}%
\caption{\label{fig1}(a) Magnetization ($M$) and d$M$/d$H$ curves at 1.4 K observed in CsCuCl$_3$ for $H$ parallel to the $c$-axis. The change in resonance frequencies ( $\Delta$$f$$_{\rm sub}$) against applied magnetic field for the $c$-axis of CsCuCl$_3$ at (b) 1.4 K and (c) various temperatures.  $\Delta$$f$$_{\rm sub}$ is the frequency difference obtained by subtracting the frequency at $T$ = 20 K as background from that at the measurement temperature. (In (a)-(c), solid and dotted lines are field ascending and descending processes, respectively.)}
\end{figure*} 

 \begin{figure}[b]
\includegraphics[keepaspectratio, scale=0.85]{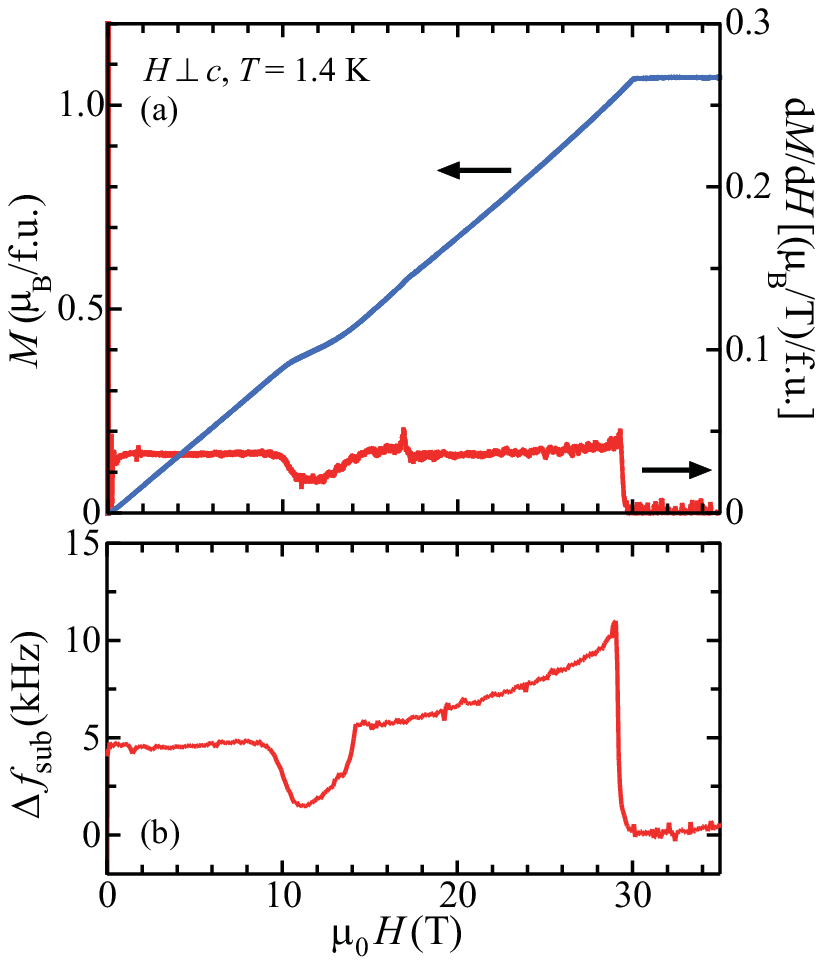}%
\caption{\label{fig2}(a) $M$ ($H$) and d$M$/d$H$ curves  for $H \perp c$ at 1.4 K. (b) The change in resonance frequency ($\Delta$$f$$_{\rm sub}$) for $H \perp c$ at 1.4 K. $\Delta$$f$$_{\rm sub}$ is the same definition as in the caption of Fig. 1. }
\end{figure}

 Figure 1(a) shows the magnetization ($M$) and its field derivative (d$M$/d$H$) of CsCuCl$_3$ for $H \parallel c$ at 1.4 K under ambient pressure. A small jump at $H_{\rm c1}$ = 12.5 T in $M$, which corresponds to a peak in d$M$/d$H$, indicates the phase transition from the umbrella to 2-1 coplanar phase. A small hysteresis in $M$ and d$M$/d$H$ was observed around $H_{\rm c1}$, indicating a first-order phase transition, and $M$ was saturated at $H_{\rm sat}$ = 30.2 T. These features were consistent with those in a previous study \cite{Nojiri}. Figure 1(b) shows the change in PDO resonance frequency versus the applied field ($\Delta$$f$$_{\rm sub}$-$H$ ) curves of CsCuCl$_3$ for $H \parallel c$ at 1.4 K under ambient pressure. The $\Delta$$f_{\rm sub}$ value was obtained by subtracting the PDO frequency at $T$ = 20 K as the background from that at each temperature. The $\Delta$$f_{\rm sub}$-$H$ curve at 1.4 K under ambient pressure indicates shoulders at $H_{\rm c1}$ and $H_{\rm sat}$, and a sharp peak at $H\rq{}_{\rm sat}$ = 29.8 T, just below $H_{\rm sat}$ (Fig.1 (b) inset). An anomaly between 18 and 24 T in the field-ascending process was observed below 4.2 K, when the applied magnetic field went beyond $H_{\rm c1}$, but the field range of the anomaly depends on the maximum applied field. The reason for this observation is currently unclear. The temperature dependence of the $\Delta$$f$$_{\rm sub}$-$H$ curve under ambient pressure is shown in Fig.1 (c). The shoulder at $H_{\rm c1}$ shifted to a lower field with increasing temperature and disappeared above $T_{\rm N}$ (=10.7 K). With increasing temperature,  the shoulder at $H_{\rm sat}$ and the sharp peak at $H\rq{}_{\rm sat}$ shifted together to a lower magnetic field, merged above 6.0 K and disappeared above $T_{\rm N}$. $H\rq{}_{\rm sat}$ may appear because of a magnetic phase transition caused by the AC response (MHz region) in the PDO measurements. In comparison, pulsed-field magnetization measurement was the low-frequency AC response at approximately 25 Hz and no sharp peak at $H\rq{}_{\rm sat}$ in d$M$/d$H$ was observed with the conventional method.\par
  In the $M$ and d$M$/d$H$ curves  for $H \perp c$ at 1.4 K under ambient pressure (Fig. 2 (a)), a plateau-like anomaly in $M$, which corresponds to a dip in d$M$/d$H$, was observed between 10 and 14 T, and the $M$ was saturated at $H_{\rm sat}$ = 29.3 T. Figure 2(b) exhibits the $\Delta$$f$$_{\rm sub}$-$H$ curve for $H \perp c$ at 1.4 K under ambient pressure.  A dip between 10 and 14 T and a peak at $H_{\rm sat}$ were observed. The sharp peak just below $H_{\rm sat}$ seen in the $\Delta$$f$$_{\rm sub}$-$H$ curves for $H \parallel c$ was not observed in the $\Delta$$f$$_{\rm sub}$-$H$ curve for $H \perp c$.

\begin{figure}[!t]
\includegraphics[keepaspectratio, scale=0.95]{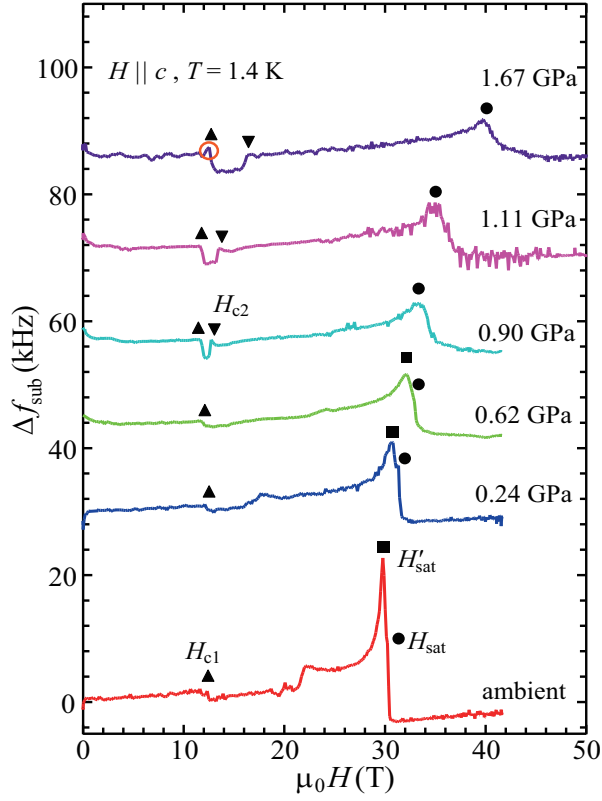}
\caption{\label{fig3} Changes in resonance frequencies $\Delta$$f$$_{\rm sub}$ ) of CsCuCl$_3$ for $H \parallel c$ at indicated pressures at 1.4 K. $\Delta$$f$$_{\rm sub}$ is the same definition as in the caption of Fig. 1. Each datum was taken from the field ascending process and shifts arbitrarily making it easier to see except for the ambient data. The small red circle at 1.67 GPa near $H_{\rm c1}$ is an anomaly which may correspond to the transition to the Y-phase. The data at ambient pressure were taken without pressure cell.}
\end{figure}
 
\begin{figure}[!t]
\includegraphics[keepaspectratio, scale=0.95]{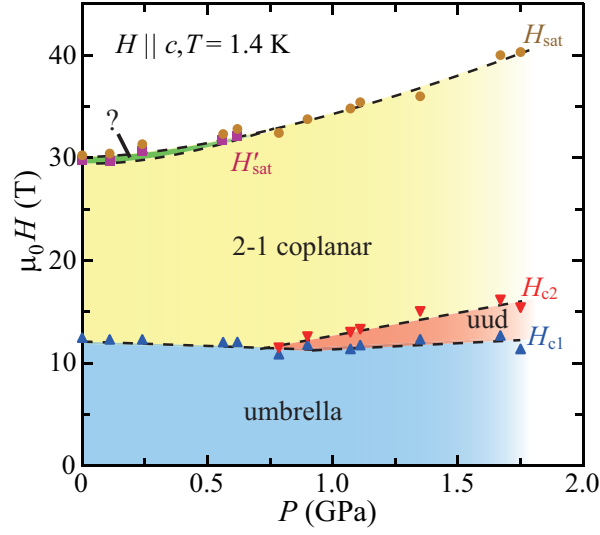}
\caption{\label{fig4}Magnetic field ($\rm \mu_0$$H$) versus pressure ($P$) phase diagram of CsCuCl$_3$ for $H \parallel c$ at 1.4 K constructed by the PDO measurements. }
\end{figure}

   \begin{figure}[b]
\includegraphics[keepaspectratio, scale=0.95]{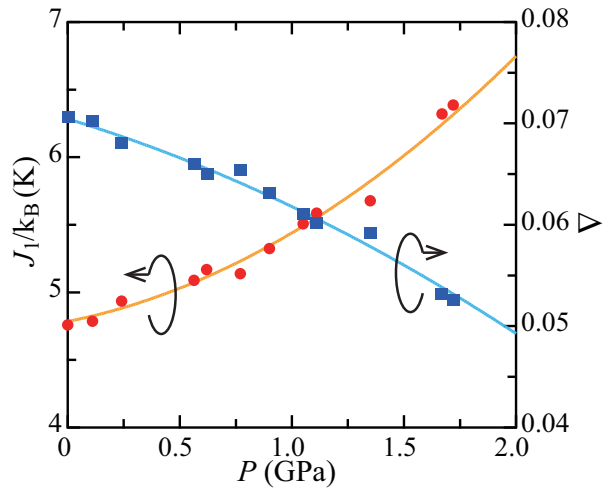}%
\caption{\label{fig5} Pressure ($P$) dependence of the interchain (in-plane) exchange constant $J_1$ and the anisotropy parameter $\Delta$, when the intrachain exchange constant $J_0$ is assumed to be constant under high pressure.}
\end{figure}

  \begin{figure*}[!t]
\includegraphics[keepaspectratio, scale=0.95]{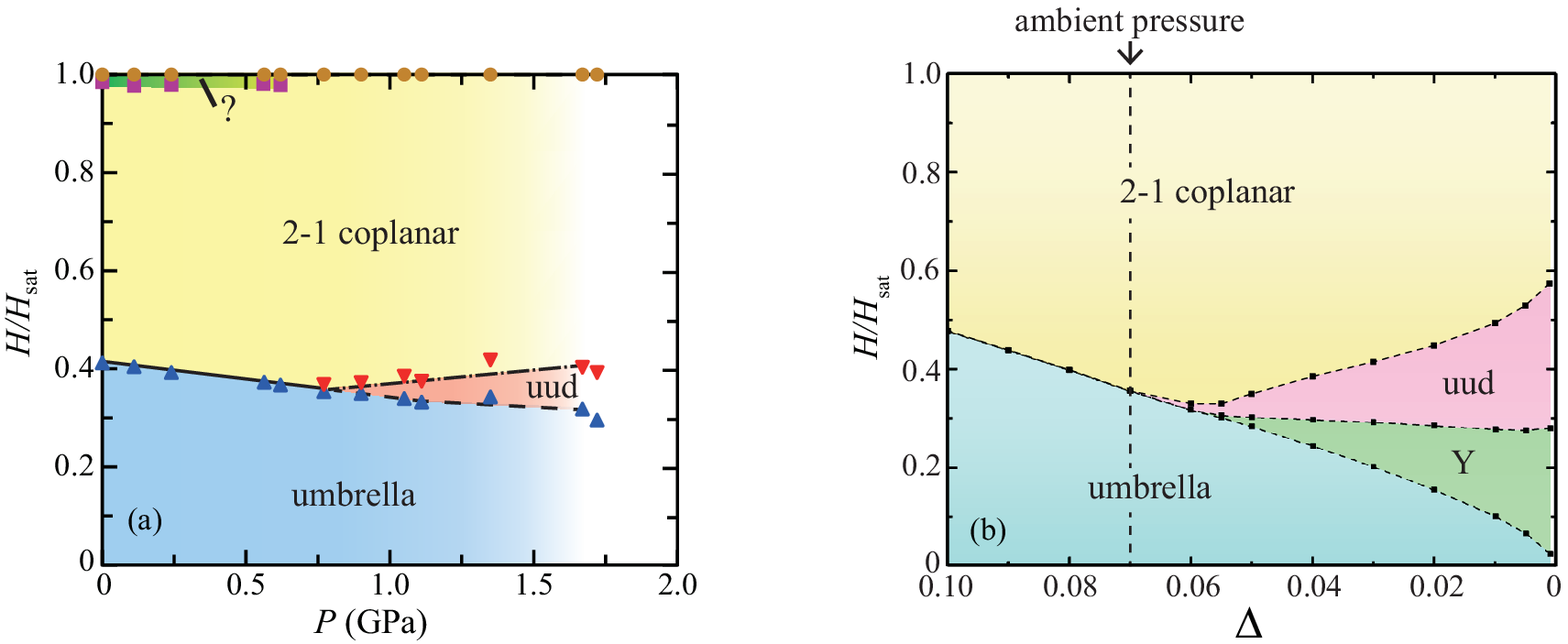}%
\caption{\label{fig6}(a) $H$/$H_{\rm sat}$ versus $P$ phase diagram of CsCuCl$_3$ for $H \parallel c$ at 1.4 K. Solid, broken, and dot-dashed lines are the phase boundary between the umbrella and 2-1 coplanar phases, between the umbrella and uud phases, and between the uud and 2-1 coplanar phases, respectively.
(b) Calculated $H$/$H_{\rm sat}$ versus $\Delta$ phase diagram constructed by reversing the horizontal axis from the phase diagram in Ref. [\onlinecite{Sera}]. The smaller $\Delta$ means the higher $P$, and $\Delta$ = 0.07 corresponds to ambient pressure. By comparison, 
$\Delta$ = 0.055 may correspond to $P$$\sim$1.7 GPa.
}
\end{figure*} 

 The pressure dependence of the $\Delta$$f$$_{\rm sub}$-$H$ curves of CsCuCl$_3$ for $H \parallel c$ at 1.4 K is shown in Fig. 3. The position of $H_{\rm c1}$ did not change significantly under $P$ $<$ 0.90 GPa. However, the shoulder at $H_{\rm c1}$ changed its structure to a dip with two shoulders when $P \geq$ 0.90 GPa. We define the low and high transition fields as $H_{\rm c1}$ and $H_{\rm c2}$, respectively. $H_{\rm c1}$ was nearly constant with increasing pressure even above 0.90 GPa. Contrarily, $H_{\rm c2}$ shifted to a higher magnetic field with increasing pressure above 0.90 GPa. The dip between $H_{\rm c1}$ and $H_{\rm c2}$ is suggested to be the uud phase reported in a previous study \cite{Nojiri}. The anomaly between 18 and 24 T at ambient pressure shifted to a lower magnetic field with increasing pressure and disappeared under $P \geq$ 0.62 GPa. The $H\rq{}_{\rm sat}$ and $H_{\rm sat}$ shifted together to a higher magnetic field with increasing pressure, and merged into one peak or broaden to form one peak at $P \geq$ 0.90 GPa. The increase in $H_{\rm sat}$ indicates an increase in the antiferromagnetic interaction in the $ab$-plane of CsCuCl$_3$, which may be caused by the shrinkage of the lattice in the $ab$-plane of CsCuCl$_3$ under pressure \cite{Christy.JPSJ.6.3125.1994}.\par
 The magnetic field ($\rm \mu_0$$H$) versus the pressure phase diagram of CsCuCl$_3$ for $H \parallel c$ at 1.4 K is shown in Fig. 4. The pressure dependence of $H_{\rm sat}$ is well-fitted by the equation $H_{\rm sat}$ = 30.2 + 2.14$P$ + 2.14$P^2$. $H_{\rm sat}$ is mainly determined by the antiferromagnetic exchange interaction in the $ab$-plane, as shown in Eq. (4). The field range of the uud phase between $H_{\rm c1}$ and $H_{\rm c2}$ expanded to a high-magnetic-field side with increasing pressure. The dashed lines extrapolated to the lower side from $H_{\rm c1}$ and $H_{\rm c2}$ intersect at approximately 0.7 GPa, which is in good agreement with the appearance of a magnetic plateau in $M$ above 0.68 GPa in statice fields observed by Sera {\it et al.} \cite{Sera}. The green magnetic phase near the saturation field in Fig. 4 is discussed in the next section.

\section{Discussion}
\subsection{Magnetic field ($\rm \mu_0$$H$) vs. pressure ($P$) phase diagram}

We determined the interchain exchange constant $J_1$ and anisotropy parameter $\Delta$ from Eqs. (3) and (4), where the ferromagnetic intrachain exchange constant $J_0$/k$_{\rm B}$ = 28 K and  $\tilde{J}_0$ = 1.012 $J_0$ were fixed under pressure, and $J_1$ ($>$ 0) is the interchain antiferromagnetic exchange constant, as defined in the introduction. Figure 5 shows the pressure dependence of $J_1$ and $\Delta$. The interchain exchange constant ($J_1$/k$_{\rm B}$ = 4.8 K at ambient pressure) was obtained from the saturation field, which is almost the same as that in Ref. [\onlinecite{Tanaka.ESR.CsCuCl3}]. $J_1$ increased with increasing pressure, whereas $\Delta$ decreased. The $J_1$/k$_{\rm B}$ and $\Delta$ were evaluated to be 6.4 K and 0.053 under the maximum pressure $P$ = 1.72 GPa, respectively. In the pressure range of this study, $\Delta$ decreased from 0.070 to 0.053 (4$\Delta$ $\ll$ 18). $J_1$ depended strongly on $H_{\rm sat}$, as shown in Eq. (4), because the change in $\Delta$ under pressure was quite small. Therefore, the pressure dependence of $J_1$ did not change much in both cases of $J_0$/k$_{\rm B}$ = constant and $J_0$/k$_{\rm B}$ = 28.45 - 10.49$P$ reported in Ref. [\onlinecite{Yamamoto.Nature.12.4263.2021}] . The magnetic-field normalized by the $H_{\rm sat}$ versus pressure phase diagram of CsCuCl$_3$ is shown in Fig. 6(a). The similar calculated phase diagram by Hosoi {\it et al.} \cite{Hosoi} is depicted in Fig. 6(b).The pressure dependences of $H_{\rm c1}$/$H_{\rm sat}$ and $H_{\rm c2}$/$H_{\rm sat}$ qualitatively agree with the calculation \cite{Hosoi}, and similar results were obtained by Yamamoto {\it et al.} \cite{Yamamoto.Nature.12.4263.2021}. The emergence of the Y coplanar phase was expected at $\Delta$ $\leq$ 0.05 and $H_{\rm c1}$/$H_{\rm sat}$ $\leq$ 0.3, as shown in Fig. 6 (b). In the $\Delta$$f$$_{\rm sub}$-$H$ curves for $H \parallel c$ at 1.4 K under $P$ = 1.67 GPa, a small peak indicated by a red circle at $H_{\rm c1}$ corresponds to the transition field from the umbrella to predicted Y coplanar phase \cite{Hosoi}.  The Y phase was expected to appear below $H_{\rm c1}$/$H_{\rm sat}$ = 0.3 in Ref. [\onlinecite{Hosoi}] (Fig. 6 (b)), but the experimentally observed $H_{\rm c1}$/$H_{\rm sat}$ was approximately 0.3 at the maximum pressure in this study. Therefore, we plan to perform PDO measurements at pressures above 1.72 GPa. A crystal structural phase transition in CsCuCl$_3$, however, was predicted at pressures between 1.65 and 3.1 GPa \cite{Christy.JPSJ.6.3125.1994}, and this transition may hinder the appearance of the Y-phase.

\subsection{The anomaly just below the saturation field}

  \begin{figure}[t]
\includegraphics[keepaspectratio, scale=0.9]{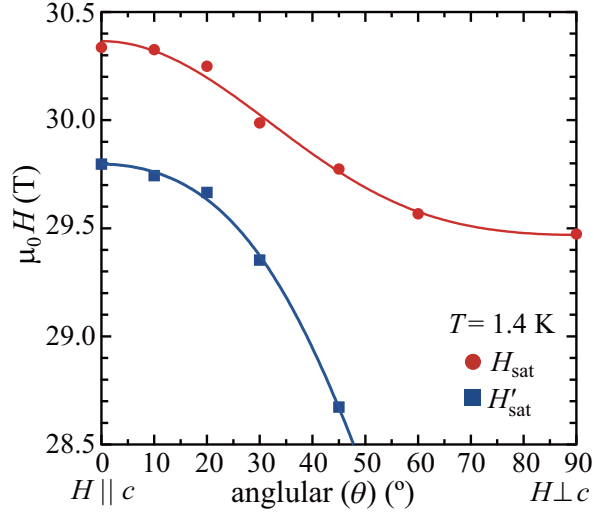}%
\caption{\label{fig7} Angular ($\theta$) dependence of $H\rq{}_{\rm sat}$ and $H_{\rm sat}$ of CsCuCl$_3$ at 1.4 K. The solid lines are the fitting curves of $H_{\rm sat}$ and $H\rq{}_{\rm sat}$ with $H_{\rm sat}$ = $A$cos2$\theta$ + $B$cos4$\theta$ + $C$ and $H\rq{}_{\rm sat}$ = $A$$\rq{}$cos2$\theta$ + $B$$\rq{}$cos4$\theta$ + $C$$\rq{}$ ($A$, $B$, $C$, $A$$\rq{}$, $B$$\rq{}$, and $C$$\rq{}$ are fitting parameters).}
\end{figure}

The peak at $H\rq{}_{\rm sat}$ observed in the $\Delta$$f$$_{\rm sub}$-$H$ curve for $H \parallel c$ did not appear in the case of  $H \perp c$. The angular dependences of $H_{\rm sat}$ and $H\rq{}_{\rm sat}$ are shown in Fig. 7, where $\theta$ is the angle between the applied magnetic field and $c$-axis. $H_{\rm sat}$ shifted gradually to a lower magnetic field with increasing. $\theta$, but $H\rq{}_{\rm sat}$ shifted abruptly to a lower magnetic field with increasing $\theta$ until $\theta$ = 45$^{\circ}$, and then disappeared at $\theta$ $>$ 45$^{\circ}$. The angle dependences of $H_{\rm sat}$ and $H\rq{}_{\rm sat}$ were fitted by the equations $H_{\rm sat}$ = $A$cos2$\theta$ + $B$cos4$\theta$ + $C$ and $H\rq{}_{sat}$ = $A$$\rq{}$cos2$\theta$ + $B$$\rq{}$cos4$\theta$ + $C$$\rq{}$($A$, $B$, $C$, $A$$\rq{}$, $B$$\rq{}$ and $C$$\rq{}$ are fitting parameters). The following parameter values were obtained from the fittings: $A$ = 0.448, $B$ = 0.078, $C$ = 29.8, $A$$\rq{}$ = 1.69, $B$$\rq{}$ = -0.280, and $C$$\rq{}$ = 28.4. In both cases, the coefficients of cos2$\theta$ were dominant, although their magnitudes were largely different.\par
 In the d$M$/d$H$ curve of the triangular lattice antiferromagnet NiBr$_2$\cite{katsumata1983NiBr2}, a sharp peak, which corresponds to the antiferro-fan transition, was observed just below the saturation field. If the peak at $H\rq{}_{\rm sat}$ indicates a magnetic transition from the 2-1 coplanar to fan phase, it should be observed in conventional magnetization measurements of CsCuCl$_3$. However, no anomaly at $H\rq{}_{\rm sat}$ was observed in the $M$ and d$M$/d$H$ curves (Fig. 1(a)). \par
Why was the $H\rq{}_{\rm sat}$ anomaly observed in the PDO measurements? In the PDO measurements, $\Delta M$/$\Delta H$ in the insulating sample CsCuCl$_3$ was taken as $\Delta$$f$ in high-frequency response (of the order of ten mega-hertz). The high-frequency response in magnetic fields may give rise to a nonlinear magnetic response owing to the third-harmonic component of $\chi_3$ in the AC magnetic susceptibility measurements ($M$ in the AC magnetic measurements is given by $M$ ($H$) = $M$$_0$ +$\chi_1$$H$ +$\chi_3$$H^3$ + $\cdot \cdot $). To clarify this interpretation, AC magnetic susceptibility measurements of CsCuCl$_3$ in high magnetic fields are required.

\section{Summary}

We experimentally performed magnetic measurements of single-crystal samples of CsCuCl$_3$ in pulsed-high magnetic fields under high pressure using the PDO technique. We succeeded in measuring the shift of the PDO frequency that corresponds to the field derivation of magnetization over the saturation field, and clarified the pressure dependence of $J_0$ and $\Delta$. The magnetic field versus pressure phase diagram of CsCuCl$_3$ constructed from the experimental results agrees qualitatively with that obtained by theoretical calculations \cite{Hosoi, Yamamoto.Nature.12.4263.2021}.The anomaly just below the saturation field for $H \parallel c$ at 1.4 - 4.2 K was observed in the PDO measurements and not in the conventional magnetization measurements. This anomaly may correspond to the transition field caused by the nonlinear AC magnetic response in PDO measurements.\\

\begin{acknowledgments}
We thank T. Sakurai, D. Yamamoto and T. Takeuchi for the fruitful discussion on the magnetism of CsCuCl$_3$ under pressure. We would like to thank Editage (www. editage.com) for English language editing. This study was financially supported by the Sasakawa Scientific Research Grant from The Japan Science Society and Osaka University Fellowship: \lq\lq{}Super Hierarchical Materials Science Program\rq\rq{}.This study was also supported by JSPS KAKENHI (Grant Nos JP17H06137, JP17K18758, JP21H01035, JP19H00648, 25220803) and JSPS Core-to-Core Program, A. Advanced Research Networks.
\end{acknowledgments}

\bibliography{reference}

\end{document}